\documentstyle[aps,12pt]{revtex}
\begin{document}
\title{Exact derivation of the Langevin and master equations for harmonic quantum
Brownian motion}
\author{Edgardo T. Garcia Alvarez and Fabi\'an H. Gaioli}
\address{Instituto de Astronom\'\i a y F\'\i sica del Espacio, \\
C.C. 67, Suc. 28, 1428 Buenos Aires, Argentina\\
Departamento de F\'\i sica, Facultad de Ciencias Exactas y Naturales,\\
Universidad de Buenos Aires, 1428 Buenos Aires, Argentina}
\maketitle

\begin{abstract}
A many particle Hamiltonian, where the interaction term conserves the number
of particles, is considered. A master equation for the populations of the
different levels is derived in an exact way. It results in a local equation
with time-dependent coefficients, which can be identified with the
transition probabilities in the golden rule approximation. A
reinterpretation of the model as a set of coupled harmonic oscillators
enables one to obtain for one of them an exact local Langevin equation, with
time-dependent coefficients.
\end{abstract}

\bigskip\ 

Pacs: 05.40.+j


\bigskip\ 

Send proof to: Fabi\'an H. Gaioli,

Instituto de Astronom\'\i a y F\'\i sica del Espacio,

C.C. 67, Suc. 28, 1428 Buenos Aires, Argentina

e-mail: gaioli@iafe.uba.ar

fax: (54-1) 786-8114

TE: (54-1) 781-6755

\newpage\ 

\section{Introduction}

The master and Langevin equations are the standard approaches to deal with
irreversible phenomena. However, it is an old problem of physics to obtain
such macroscopic irreversible equations departing from the reversible
microscopic laws without appealing, {\it a priori}, to approximations or
extra-dynamical hypotheses.

The simplest systems for which the origin of irreversibility can be studied
on a microscopic basis are the harmonic linear ones \cite{varios,ggg}, since
one is able to reduce the Hamiltonian to normal modes, for which the
dynamical evolution becomes trivial. With this purpose we consider
many particle systems with Hamiltonians of the form 
\begin{equation}
H=\sum_{n=0}^N\sum_{m=0}^N\left\langle \psi _n|h|\psi _m\right\rangle
b_n^{\dagger }b_m,  \label{H}
\end{equation}
which conserve the total number of quanta. For this kind of system it can be
proved by a straightforward calculation that the time evolution of the
creation operators is given by\footnote{%
The meaning of Eq. (\ref{bAmn}) is clarified applying to both members of the
equality the vacuum state. This means that the operator $b_n^{\dagger }(t)$
creates states $\left| \psi _n(t)\right\rangle .$}

\begin{equation}
b_n^{\dagger }(t)=e^{-iHt}b_n^{\dagger
}(0)e^{iHt}=\sum_mA_{nm}(t)b_m^{\dagger }(0),  \label{bAmn}
\end{equation}
being $A_{nm}(t)=\left\langle \psi _m|e^{-iht}|\psi _n\right\rangle $ the
transition amplitude between one-particle states.

From Eq. (\ref{bAmn}) 
the exact
``Langevin'' and ``master'' equations can be derived. 
This work is devoted to study these
equations in order to trace the origin of macroscopic irreversibility. In
this route we show, from the reversible evolution of the system, which are the
simplest, {\it a posteriori,} extra-dynamical hypotheses needed to
understand the irreversible behavior.

\section{The master equation}

For systems described by a Hamiltonian such as those of Eq. (\ref{H}) we can
obtain an exact master equation for the mean occupation number $\left\langle
N_n\right\rangle =\left\langle b_n^{\dagger }b_n\right\rangle ,$
corresponding to the state $\left| \psi _n\right\rangle $ of the
one-particle Hamiltonian $h$, namely

\begin{equation}
\frac{d\left\langle N_n(t)\right\rangle }{dt}=\sum_kW_{nk}(t)\left\langle
N_k(t)\right\rangle ,  \label{em1}
\end{equation}
or equivalently in the form of a kinetic balance equation 
\begin{equation}
\frac{d\left\langle N_n(t)\right\rangle }{dt}=\sum_{m\neq n}\left[
W_{nm}(t)\left\langle N_m(t)\right\rangle -W_{mn}(t)\left\langle
N_n(t)\right\rangle \right] .  \label{em2}
\end{equation}
The time-dependent (non-symmetrical) coefficients $W_{nm}(t)$ are given by 
\begin{equation}
W_{nk}(t)=\sum_m\stackrel{.}{P}_{nm}(t)P_{mk}^{-1}(t),  \label{coef}
\end{equation}
where $P_{nm}(t)=\left| A_{nm}(t)\right| ^2$ is the probability to find a
quantum in the state $\left| \psi _m\right\rangle $ at time $t$ if it was in
the state $\left| \psi _n\right\rangle $ at $t=0.$ 

We can prove Eq. (\ref
{em1}) as follows: First, from (\ref{bAmn}), it can be checked that\footnote{%
Eq. (\ref{solM}) has a simple interpretation. The mean number of quanta at
the level $n$ at a given time $t$ can be obtained as a sum of the initial
populations of different levels times the transition probabilities at $t$
from these levels (including the own $n$) to the level $n$.}

\begin{equation}
\left\langle N_n(t)\right\rangle =\sum_mP_{nm}(t)\left\langle
N_m(0)\right\rangle ,  \label{solM}
\end{equation}
provided we assumed a privileged initial condition which represents the
absence of initial correlations\footnote{%
The last hypothesis has the same status as the random phase aproximation.}

\begin{equation}
\left\langle b_m^{\dagger }(0)b_n(0)\right\rangle =\delta _{mn}\left\langle
N_m(0)\right\rangle .  \label{correla}
\end{equation}

Finally, as is well known, departing from the solution we can form a
differential equation by derivation and eliminating the integration
constants. That is, derivating (\ref{solM}) with respect to time and
inverting this linear system to eliminate $\left\langle N_m(0)\right\rangle
, $ we finally obtain (\ref{em1}).

\section{The Langevin equation}

Until now the basis $\left\{ \left| \psi _n\right\rangle \right\}
_{n=0,...,N}$ is any complete set of the one-particle space. In this section
we consider the case in which the basis $\left\{ \left| \psi _n\right\rangle
\right\} _{n=0,...,N}$ diagonalizes the unperturbed one-particle Hamiltonian 
$h_0$ ($h=h_0+v),$ such that $H=H_0+V$ can be thought as a set of
interacting harmonic oscillators. $H_0$ represents the set of uncoupled
oscillators and $V$ is a linear interaction among them. Splitting the
summation in such a way that one oscillator is identified as a Brownian
particle and the rest as a bosonic reservoir, redefining the notation by $%
\left\{ \left| \psi _0\right\rangle ,\left| \psi _n\right\rangle \right\}
_{n=1,...,N}=\left\{ \left| \Omega \right\rangle ,\left| \omega
_n\right\rangle \right\} _{n=1,...,N},$ $b_0=B,$ we have $\left( \hbar
=1\right) $ 
\begin{equation}
H_0=\Omega B^{\dagger }B+\sum_{n=1}^N\omega _nb_n^{\dagger }b_n,  \label{Ho}
\end{equation}
and 
\begin{eqnarray}
V &=&\left\langle \Omega |v|\Omega \right\rangle B^{\dagger
}B+\sum_{n=1}^N\sum_{m=1}^N\left\langle \omega _n|v|\omega _m\right\rangle
b_n^{\dagger }b_m+  \nonumber \\
&&\ \ \ \ \ \sum_{n=1}^N\left( \left\langle \omega _n|v|\Omega \right\rangle
b_n^{\dagger }B+\left\langle \Omega |v|\omega _n\right\rangle B^{\dagger
}b_n\right) .  \label{V}
\end{eqnarray}
In this case we can derive a generalized form of the Langevin equation for
the position operator $X=\frac 1{\sqrt{2M\Omega }}\left( B+B^{\dagger
}\right) $ of the oscillator with frequency $\Omega $ departing from the
exact solution (\ref{bAmn}) with $n=0$, namely

\begin{equation}
X(t)=a(t)X(0)+b(t)\frac{P(0)}{M\Omega }+f(t),  \label{Xt}
\end{equation}
\[
A_{\Omega \Omega }(t)=a(t)+ib(t),\hspace{0.3in}f(t)=\frac 1{\sqrt{2M\Omega }}%
\sum_{m=1}^N\left[ A_{\Omega m}(t)b_m^{\dagger }(0)+h.c.\right] . 
\]
As in this case we have two constants of integration and a particular
solution $f(t),$ $X(t)$ satisfies a second-order differential equation like 
\begin{equation}
\stackrel{..}{X}(t)+\Omega ^2(t)X(t)+\Gamma (t)\stackrel{.}{X}(t)=F(t),
\label{el}
\end{equation}
with an inhomogeneous term given by 
\[
F(t)=\stackrel{..}{f}(t)+\Omega ^2(t)f(t)+\Gamma (t)\stackrel{.}{f}(t), 
\]
which represents the analogue of the stochastic acceleration in the standard
Langevin equation.\footnote{%
However, its deterministic character is obvious from its definition.} The
unknown coefficients $\Omega ^2(t)$ and $\Gamma (t),$ are the analogues of
the time-dependent square frequency and damping factor, respectively.
They can be easily determined by solving the linear system which results in
replacing the two independent solutions of the homogeneous equations:

\begin{eqnarray*}
\stackrel{..}{a}(t)+\Omega ^2(t)a(t)+\Gamma (t)\stackrel{.}{a}(t) &=&0, \\
&& \\
\stackrel{..}{b}(t)+\Omega ^2(t)b(t)+\Gamma (t)\stackrel{.}{b}(t) &=&0.
\end{eqnarray*}
That is

\begin{equation}
\Omega ^2(t)=\frac{\stackrel{.}{A_{\Omega \Omega }}\stackrel{..}{A_{\Omega
\Omega }^{*}}-\stackrel{.}{A_{\Omega \Omega }^{*}}\stackrel{..}{A_{\Omega
\Omega }}}{A_{\Omega \Omega }\stackrel{.}{A_{\Omega \Omega }^{*}}-A_{\Omega
\Omega }^{*}\stackrel{.}{A_{\Omega \Omega }}},\hspace{0.3in}\Gamma (t)=-%
\frac{A_{\Omega \Omega }\stackrel{..}{A_{\Omega \Omega }^{*}}-A_{\Omega
\Omega }^{*}\stackrel{..}{A_{\Omega \Omega }}}{A_{\Omega \Omega }\stackrel{.%
}{A_{\Omega \Omega }^{*}}-A_{\Omega \Omega }^{*}\stackrel{.}{A_{\Omega
\Omega }}}.  \label{coel}
\end{equation}

\section{Perturbative calculations}

The coefficients of the master equation are in general time-dependent. We
can evaluate the transition probabilities involved in these coefficients
using time-dependent perturbation theory. The Fermi golden rule gives $%
P_{nm}=\delta _{nm}+\Gamma _{nm}t,$ where

\[
\Gamma _{nm}=2\pi \left| \left\langle \psi _n\right| v\left| \psi
_m\right\rangle \right| ^2\delta _t(\omega _n-\omega _m),\hspace{0.2in}\text{%
for }n\neq m, 
\]

\[
\Gamma _{nn}=-\sum_{n\neq m}2\pi \left| \left\langle \psi _n\right| v\left|
\psi _m\right\rangle \right| ^2\delta _t(\omega _n-\omega _m), 
\]
being $\delta _t(\alpha )$ $=\frac{\sin ^2\alpha t}{\pi \alpha t}$ an
approximating of the Dirac delta for very long times. Using the last
expressions for calculating coefficients $W_{nm}(t),$ neglecting higher
order terms in the perturbation, we have the time-independent symmetrical
coefficients $W_{nm}(t)=\sum_k\Gamma _{nk}(\delta _{km}-\Gamma
_{km}t)=\Gamma _{nm},$ in agreement with standard results. In the case
of the coefficients of the Langevin equation we must evaluate up to the
second order the survival amplitude of the state $\left| \Omega
\right\rangle $. For very long times an exponential contribution dominates
its time evolution, $A_{\Omega \Omega }(t)=e^{-i(\Omega +\delta \Omega -i%
\frac \gamma 2)t},$ with a frequency shift and a damping factor given by 
\[
\delta \Omega =\left\langle \Omega \right| v\left| \Omega \right\rangle +%
{\rm P}\sum_{k=1}^N\frac{\left| \left\langle \omega _k\right| v\left| \Omega
\right\rangle \right| ^2}{\Omega -\omega _k},\hspace{0.3in}\gamma =2\pi
\left| \left\langle \Omega \right| v\left| \Omega \right\rangle \right| ^2. 
\]
A straightforward calculation shows that in the exponential decay regime the
coefficients of Eq. (\ref{coel}) are given by $\Omega (t)=\Omega +\delta
\Omega $ and $\Gamma (t)=\gamma $ .

\section{Conclusions and further remarks}

We have shown that from reversible quantum mechanical laws we can obtain
equations of motion as if they represented stochastic processes. However,
our equations (\ref{em1}) and (\ref{el}) describe the exact dynamical
evolution of the system and in this sense they are equivalent to the
Schr\"odinger and/or Heisenberg equations. We also stress the fact that Eqs.
(\ref{em1}) and (\ref{el}) are local in time in contrast with the standard
non-Markovian master and Langevin equations. In our case all memory effects
are reduced to the knowledge of the initial conditions. The time-dependent
coefficients of Eqs. (\ref{em1}) and (\ref{el}) are uniquely determined for
the amplitudes $A_{nm}(t)$ of the one-particle sector. Evaluating these
amplitudes through perturbation theory, retaining up to the second order in
the Dyson expansion, we retrieve the standard irreversible equations with
time-independent coefficients. In Ref. \cite{ggg} we have considered a
particular model in which there is not interaction among bath oscillators ($%
\left\langle \omega _n\right| v\left| \omega _m\right\rangle =0$)$.$ This
choice allows us to reduce $H$ to normal modes (a set of uncoupled harmonic
oscillators), which in the one-particle sector means that $h$ is diagonal: $%
h=\sum_{\nu =0}^N\alpha _\nu \left| \alpha _\nu \right\rangle \left\langle
\alpha _\nu \right| .$ In this case the transition amplitudes can be
analytically obtained as

\begin{equation}
A_{nm}(t)=\sum_{\nu =0}^Ne^{-i\alpha _\nu t}\left\langle \psi _m|\alpha _\nu
\right\rangle \left\langle \alpha _\nu |\psi _n\right\rangle ,  \label{ampl}
\end{equation}
which will allow us to extend the previous analysis beyond the perturbative 
calculations and
determine its range of applicability in a further work. 

\smallskip\ 

We thank Leopoldo Garc\'\i a-Col\'\i n for fruitful discussions.

\end{document}